\pdfoutput=1
\newcommand*{\ATLASLATEXPATH}{./}
\documentclass[USenglish, texlive=2016, cernpreprint]{\ATLASLATEXPATH atlasdoc}

\usepackage[subfigure]{\ATLASLATEXPATH atlaspackage}
\usepackage{\ATLASLATEXPATH atlasbiblatex}

\usepackage[process]{\ATLASLATEXPATH atlasphysics}

\usepackage{EXOT-2017-06-PAPER-defs}

\usepackage[colorinlistoftodos,prependcaption,textsize=tiny]{todonotes}
\usepackage{siunitx}
\usepackage[inline,shortlabels]{enumitem}
\usepackage{longtable}

\AtlasTitle{Search for High-Mass Resonances Decaying to $\tau\nu$ in $pp$ Collisions at $\sqrt{s}=\SI{13}{\TeV}$ with the ATLAS Detector}

\AtlasAbstract{%

A search for high-mass resonances decaying to $\tau\nu$ using proton--proton
collisions at \mbox{$\sqrt{s}=\SI{13}{\TeV}$} produced by the Large Hadron
Collider is presented.  Only $\tau$-lepton decays with hadrons in the 
final state are considered.  The data were recorded with the ATLAS detector and correspond to
an integrated luminosity of \SI{36.1}{\ifb}.  No statistically significant
excess above the standard model expectation is observed; model-independent
upper limits are set on the visible $\tau\nu$ production cross section.
Heavy $W^{\prime}$ bosons with masses less than \SI{3.7}{\TeV} in the sequential
standard model and masses less than \mbox{2.2--\SI{3.8}{\TeV}} depending on
the coupling in the non-universal $G(221)$ model are excluded at the 95\%
credibility level. 

}

\author{The ATLAS Collaboration}

\AtlasRefCode{EXOT-2017-06}

\PreprintIdNumber{CERN-EP-2017-341}

\AtlasJournal{Phys. Rev. Lett.}
\AtlasJournalRef{\PRL120 (2018) 161802}
\AtlasDOI{10.1103/PhysRevLett.120.161802}

\hypersetup{pdftitle={ATLAS document},pdfauthor={The ATLAS Collaboration}}

\begin{document}

\maketitle

Heavy charged gauge bosons (\Wprime) appear frequently in theories of physics
beyond the standard model (SM).  They are often assumed to obey lepton
universality, such as in the sequential standard model
(SSM)~\cite{Altarelli1989}, which predicts a \WprimeSSM boson with couplings
identical to those of the SM $W$ boson.  However, this assumption is not
required. In particular, models in which the \Wprime boson couples
preferentially to third-generation fermions may be linked to the high mass of
the top quark~\cite{sfm, TopFlavor, G221, Family221} or to recent indications
of lepton flavor universality violation in $B$ meson decays~\cite{BPhys1,
BPhys2}.
An example is the non-universal $G(221)$ model (NU)~\cite{G221, Family221},
which exhibits a $SU(2)_l \times SU(2)_h \times U(1)$ gauge symmetry, where
$SU(2)_l$ couples to \emph{light} fermions (first two generations), $SU(2)_h$
couples to \emph{heavy} fermions (third generation) and $\phinu$ is the mixing
angle between them.  The model predicts \WprimeSFM and \ZprimeSFM bosons which
are approximately degenerate in mass and couple only to left-handed fermions.
At leading order and neglecting sign, the \WprimeSFM couplings to heavy (light)
fermions are scaled by $\cot\phinu$ ($\tan\phinu$) relative to those of \WprimeSSM.
Thus $\cot\phinu>1$ corresponds to enhanced couplings to tau leptons while
$\cot\phinu=1$ yields \WprimeSFM couplings identical to those of \WprimeSSM. For
\ZprimeSFM, the coupling to heavy (light) fermions is given by $g\cot\phinu$
($g\tan\phinu$), where $g$ is the SM weak coupling constant.  At high values of
$\cot\phinu$, the branching fraction of \WprimeSFM to a tau lepton ($\tau$) and a
neutrino ($\nu$) approaches 26\%.
 
In this Letter, a search for high-mass resonances (0.5--\SI{5}{\TeV}) decaying 
to $\tau\nu$ using proton--proton ($pp$) collisions at a
center-of-mass energy of $\sqrt{s}=\SI{13}{\TeV}$ produced by the Large Hadron
Collider (LHC) is presented. The data were recorded with the ATLAS detector and
correspond to an integrated luminosity of \SI{36.1}{\ifb}.  Only $\tau$ 
decays with hadrons in the final state are considered; these account for 65\% of the
total $\tau$ branching fraction.  A counting experiment is performed from
events that pass a high transverse-mass threshold, optimized separately for
each of the signal mass hypotheses.

A direct search for high-mass resonances decaying to \taunu has been performed
by the CMS Collaboration using \SI{19.7}{\ifb} of integrated luminosity at
$\sqrt{s}=\SI{8}{\TeV}$~\cite{CMS-EXO-12-011}. The search excludes \WprimeSSM
with a mass below \SI{2.7}{\TeV} at the 95\% credibility level and \WprimeSFM
with a mass below 2.7--\SI{2.0}{\TeV} for $\cot\phinu$ in the range 1.0--5.5. 
The most stringent limit on \WprimeSSM from searches in the \enu and \munu
final states is \SI{5.1}{\TeV} from ATLAS~\cite{EXOT-2016-06} using
\SI{36.1}{\ifb} of integrated luminosity at $\sqrt{s} = \SI{13}{\TeV}$.

The ATLAS experiment is a multipurpose particle detector with a
forward-backward symmetric cylindrical geometry~\cite{PERF-2007-01, ATLASFootnote}. It
consists of an inner detector for charged-particle tracking in the pseudorapidity 
region $|\eta|<2.5$, electromagnetic and hadronic calorimeters that provide energy
measurements up to $|\eta|=4.9$, and a muon spectrometer that covers $|\eta| <
2.7$. A two-level trigger system is used to select events
\cite{TRIG-2016-01-bugfix}. 

Hadronic $\tau$ decays are composed of a neutrino and a set of visible decay
products (\tauhadvis), typically one or three charged pions and up to two
neutral pions. The reconstruction of the visible decay
products~\cite{PERF-2013-06} is seeded by jets reconstructed from topological
clusters of energy depositions \cite{Aad:2016upy} in the calorimeter. The
\tauhadvis candidates must have a transverse momentum $\pt > \SI{50}{\GeV}$,
$|\eta|<2.4$ (excluding $1.37<|\eta|<1.52$), one or three associated tracks and
an electric charge of $\pm1$.  Only the candidate with the highest \pt in each
event is selected.  Hadronic $\tau$ decays are identified using boosted
decision trees that exploit calorimetric shower shape and tracking
information~\cite{ATL-PHYS-PUB-2015-045-fixed, ATLAS-CONF-2017-029}.  Loose
criteria are used, which offer adequate rejection against quark- and
gluon-initiated jets. Very loose criteria, with about one quarter of the
rejection power, are used to create control regions.  An additional dedicated
veto is used to reduce the number of electrons misidentified as \tauhadvis.
The total efficiency for \tauhadvis is $\sim$60\% at $\pt=\SI{100}{\GeV}$ and
decreases to $\sim$30\% at $\pt=\SI{2}{\TeV}$, where the large boost and
collimation of the decay products causes inefficiencies in the track
reconstruction and association.

Events containing electron or muon candidates are rejected.  Electron
candidates~\cite{PERF-2013-03,PERF-2013-05,ATLAS-CONF-2016-024} must have $\pt
> \SI{20}{\GeV}$, $|\eta|<2.47$ (excluding $1.37<|\eta|<1.52$) and must pass a
loose likelihood-based identification selection.  Muon
candidates~\cite{PERF-2015-10} are required to have $\pt>\SI{20}{\GeV}$,
$|\eta|<2.5$ and to pass a very loose muon identification requirement.  The
missing transverse momentum, with magnitude \met, is calculated as the negative
vectorial sum of the \pt of all reconstructed and calibrated \tauhadvis
candidates and jets~\cite{PERF-2016-04, ATL-PHYS-PUB-2015-027,
ATL-PHYS-PUB-2015-023}.  A correction that accounts for momentum not associated
with these reconstructed objects is calculated using inner-detector tracks that
originate from the hard-scattering vertex~\cite{ATL-PHYS-PUB-2015-023}. The correction contributes 
no more than 5\% on average in signal events. 

Events are selected by triggers that require \met above thresholds of 70, 90 or
\SI{110}{\GeV} depending on the data-taking period.  To minimize uncertainties
in the trigger efficiency, the offline reconstructed \met is required to be at
least \SI{150}{\GeV}.  At this threshold the trigger efficiency is 80\% and increases
to more than 98\% above $\SI{250}{\GeV}$. This behavior is determined by the 
\met resolution of the trigger, which is lower than in the offline reconstruction. The events must satisfy criteria
designed to reduce backgrounds from cosmic rays, single-beam-induced events and
calorimeter noise~\cite{ATLAS-CONF-2015-029} and they must contain a loose
\tauhadvis candidate. To further suppress single-beam-induced background, the
\tauhadvis must have at least one associated track with $\pt>\SI{10}{\GeV}$.
The multijet background is further suppressed by requiring that the \tauhadvis \pt
and the \met are balanced: $0.7 < \taupt / \met < 1.3$.  The azimuthal angle
between the \tauhadvis and the missing momentum, $\Delta\phi$, is required to
be larger than 2.4. Finally, thresholds ranging from \SI{0.25}{\TeV} to \SI{1.8}{\TeV} in
steps of \SI{0.05}{\TeV} are placed on the transverse mass, \mt, where $\mt^2
\equiv 2\taupt\met(1-\cos\Delta\phi)$.

The background is divided into events where the selected \tauhadvis originates
from a quark- or gluon-initiated jet (jet background) and those where it
does not (non-jet background). The jet background originates primarily from $\WZjets$ and \multijet
production and is estimated using a data-driven technique. The non-jet 
background is estimated using simulation and originates primarily from \Wtaunu
production with additional minor contributions from \WZg, \ttbar, \singletop
and \diboson ($WW$, $WZ$ and $ZZ$) production (collectively called
\emph{others}).

The event generators and other software packages used to produce the simulated samples are summarized in
Table~\ref{tab:samples}. The \WZg sample is artificially enhanced in 
high-mass events to improve statistical coverage in the scanned mass range. 
Particle interactions with the ATLAS detector are simulated with 
\GEANT{}4~\cite{Geant4,SOFT-2010-01} and contributions from 
additional $pp$ interactions (pileup) are simulated using 
\PYTHIAV{8.186} and the MSTW2008LO parton distribution function (PDF) set~\cite{Martin:2009iq}.
Finally, the
simulated events are processed through the same reconstruction software as the
data. Corrections are applied to account for mismodeling of the momentum scales
and resolutions of reconstructed objects, the \tauhadvis reconstruction and
identification efficiency, the electron to \tauhadvis misidentification rate
and the \met trigger efficiency.

\begin{table}[tp]
    \centering
    \scriptsize
    \caption{
        The event generators and other software packages used to generate the matrix-element process and model 
        non-perturbative effects in the simulated event samples. 
        The top-quark mass is set to \SI{172.5}{\GeV}.
    }
    \label{tab:samples}
    \begin{tabular}{llll}

\toprule
Process                           & Matrix Element          & Non-perturbative & Refs.   \\ 
\midrule
\WZg         & \POWHEGBOXV{2}, CT10, \PHOTOSppV{3.52} 
             & \PYTHIAV{8.186}, AZNLO, CTEQ6L1, \EVTGENV{1.2.0}
             & \cite{POWHEG, PowhegWZ, PowhegWZjets, CT10, Davidson:2010ew, PYTHIA82, ATLASZPTTune, CTEQ6L1, EvtGen}\\
\ttbar       & \POWHEGBOXV{2}, CT10
             & \PYTHIAV{6.428}, P2012, CTEQ6L1, \EVTGENV{1.2.0}
             & \cite{Powhegttbar, Pythia, Skands:2010ak} \\
Single top   & \POWHEGBOXV{1}, CT10f4, \MADSPIN 
             & \PYTHIAV{6.428}, P2012, CTEQ6L1, \EVTGENV{1.2.0} 
             & \cite{PowhegSingleTopST, PowhegSingleTopT4F, PowhegSingleTopW, Artoisenet:2012st} \\
Diboson      & \SHERPAV{2.1.1}, CT10
             & \SHERPAV{2.1.1} 
             & \cite{SHERPA, Gleisberg:2008fv, Cascioli:2011va, Schumann:2007mg, Hoeche:2012yf} \\
\bottomrule
    \end{tabular}
\end{table}

The simulated samples are normalized using the integrated luminosity of the
collected dataset and their theoretical cross sections.  The \WZg 
cross sections are calculated as a function of the boson mass at next-to-next-to-leading
order (NNLO)~\cite{Anastasiou:2003ds} using the CT14NNLO PDF set, including 
electroweak corrections at next-to-leading order (NLO)~\cite{Bondarenko:2013nu}
using the MRST2004QED PDF set~\cite{Martin:2004dh}. Uncertainties are taken
from Ref.~\cite{EXOT-2015-07} and include variations of the PDF sets, scale,
\alphas, beam energy and electroweak corrections. 
The variations amount to a $\sim$$5\%$ total uncertainty in the \WZg
cross section at low mass, increasing to 34\% at \SI{2}{\TeV}.  The
\ttbar and \singletop production cross sections are calculated to at
least NLO with an uncertainty of 3--6\%~\cite{Czakon:2011xx, Kant:2014oha,
Aliev:2010zk, Kidonakis:2010ux}.  The diboson cross sections are calculated to
NLO with an uncertainty of $10\%$~\cite{Campbell:2011bn,SHERPA}.

The simulated samples are affected by uncertainties associated with the
generation of the events, the detector simulation and the determination of the
integrated luminosity.  Uncertainties related to the modeling of the hard
scatter, radiation and fragmentation are at most 2\% of the total background
estimate.  Uncertainties in the detector simulation manifest themselves through
the efficiency of reconstruction, identification and triggering algorithms, and
through particle energy scales and resolutions.  The effects of energy
uncertainties are propagated to \met.  The uncertainty in the \tauhadvis
identification efficiency is 5--6\%, as determined from measurements of
\Ztautau events.  An additional uncertainty that increases by 20--25\% per
\TeV{} is assigned to \tauhadvis candidates with $\pt>\SI{150}{\GeV}$ in accord with 
studies of high-\pt jets~\cite{HIGG-2016-12}. The uncertainty in the \tauhadvis
energy scale is 2--3\%.  The probability for electrons to be misidentified as
\tauhadvis is measured with a precision of 3--14\%~\cite{ATLAS-CONF-2017-029}.
The uncertainty in the \met trigger efficiency is negligible for
$\met>\SI{300}{\GeV}$ and can be as large as 10\% for $\met<\SI{300}{\GeV}$.  Uncertainties associated
with reconstructed electrons, muons and jets are found to have a very small
impact.  The uncertainty in the combined 2015+2016 integrated luminosity is
2.1\%, derived following a methodology similar to that used in Ref.~\cite{DAPR-2013-01}, and
has a minor impact.  The uncertainty related to the simulation of pileup is $\sim$1\%. 

The \Wprime{} signal events are modeled by reweighting the $W$ sample using
a leading-order matrix-element calculation. Electroweak corrections for the
$W$ cross section and interference between $W$ and \Wprime are not included
as they are model dependent.  Uncertainties in the \Wprime cross section are
estimated in the same way as for $W$ bosons.  They are not included in the fitting procedure used
to extract experimental cross-section limits, but are instead included when
overlaying predicted model cross sections. Uncertainties in the \Wprime 
acceptance due to PDF, scale, and \alphas variations are negligible. In the NU model, the total decay
width increases to 35\% of the pole mass for large values of $\cot\phinu$, which
decreases the signal acceptance as more events are produced at low mass. Decays to $WZ$ and $Wh$ are not considered in
the calculation of the total \WprimeSFM decay width as their impact is small 
($<7\%$) and model dependent. Values of $\cot\phinu>5.5$ are not considered 
as the model is non-perturbative in this range. 

The jet background contribution is estimated using events in three control
regions (CR1, CR2 and CR3). The events must pass the selection for the signal
region, except in CR1 and CR3 they must fail loose but pass very loose
\tauhadvis identification and in CR2 and CR3 they must have
$\met<\SI{100}{\GeV}$ and the requirement on $\taupt/\met$ is removed.  The
low-\met requirement yields high multijet purity in CR2 and CR3, while the very
loose identification preferentially rejects gluon-initiated jets over
quark-initiated jets. This produces a similar fraction of quark-initiated jets in
all control regions, which ensures minimal correlation between the identification and
\met.  The estimated jet contribution is defined as:
$N_\text{jet} = N_\text{CR1} N_\text{CR2} / N_\text{CR3}$.  The non-jet
contamination in CR1 (10\%), CR2 (3.7\%) and CR3 (0.5\%) is subtracted using
simulation.  The transfer factor, $N_\text{CR2} / N_\text{CR3}$, is
parameterized in \tauhadvis \pt and track multiplicity and is in the range
\mbox{0.4--0.7} \mbox{(0.15--0.3)} for 1-track (3-track) \tauhadvis. 
Systematic uncertainties are assigned to account for any residual correlation between
the transfer factor and the \met and $\taupt/\met$ selection criteria, which would arise 
if the jet composition was different in CR1 and CR3.
They are evaluated by repeating the jet estimate with the
following modified control region definitions: (a) altered very loose
\tauhadvis identification criteria, (b) modified \met and $\taupt/\met$
selection, and (c) CR2 and CR3 replaced by alternative control regions rich in
\Wmunujets events. The corresponding variations define the dominant uncertainty
in the jet background contribution, which ranges from 20\% at
$\mt=\SI{0.2}{\TeV}$ to $^{+200\%}_{-60\%}$ at $\mt=\SI{2}{\TeV}$, where the
jet background is subdominant. The uncertainty due to the subtraction of
non-jet contamination in the control regions is negligible.

To reduce the impact of statistical fluctuations in the jet background
estimate, a function $f(\mt)=\mt^{a + b\log\mt}$, where $a$ and $b$ are free
parameters, is fitted to the estimate in the range $400 < \mt < \SI{800}{\GeV}$
and is used to evaluate the jet background in the range $\mt>\SI{500}{\GeV}$.
The impact of altering the fit range leads to an uncertainty that increases
with \mt, reaching 50\% at $\mt=\SI{2}{\TeV}$. The statistical uncertainty from the
control regions is propagated using pseudo-experiments and also reaches 50\% at
$\mt=\SI{2}{\TeV}$.

\begin{figure}[tp]
    \centering
    \includegraphics[width=0.49\textwidth]{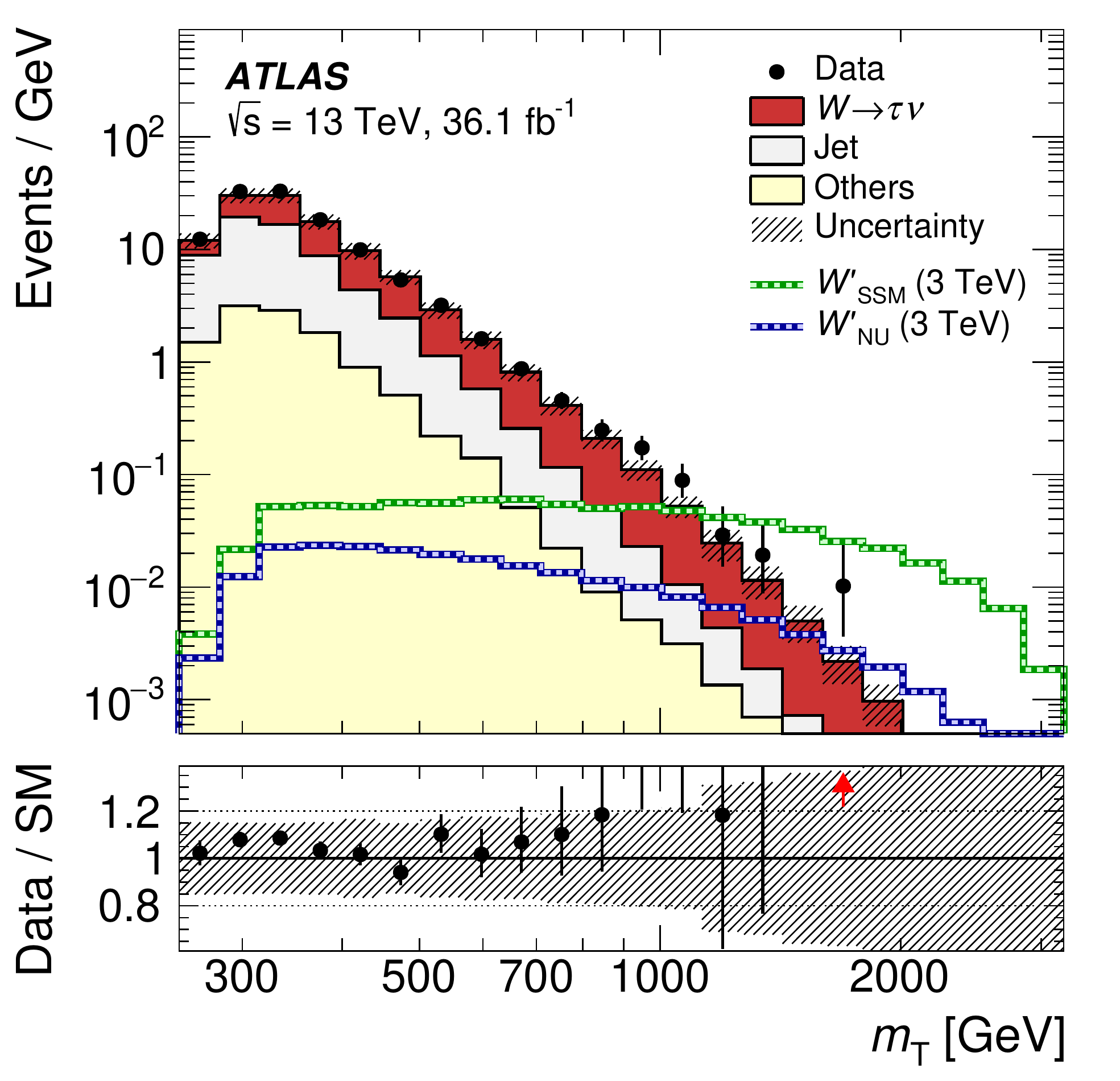}
    \caption{
    Transverse mass distribution after the event selection.
    The total impact of the statistical and systematic uncertainties on the 
    SM background is depicted by the hatched area.
    The ratio of the data to the estimated SM background is shown in the lower panel. 
    The prediction for \WprimeSSM and \WprimeSFM ($\cot\phinu=5.5$) bosons with masses of \SI{3}{\TeV} are superimposed.
    }
    \label{fig:finalmass}
\end{figure}

\begin{figure}[tp]
    \centering
    \includegraphics[width=0.49\textwidth]{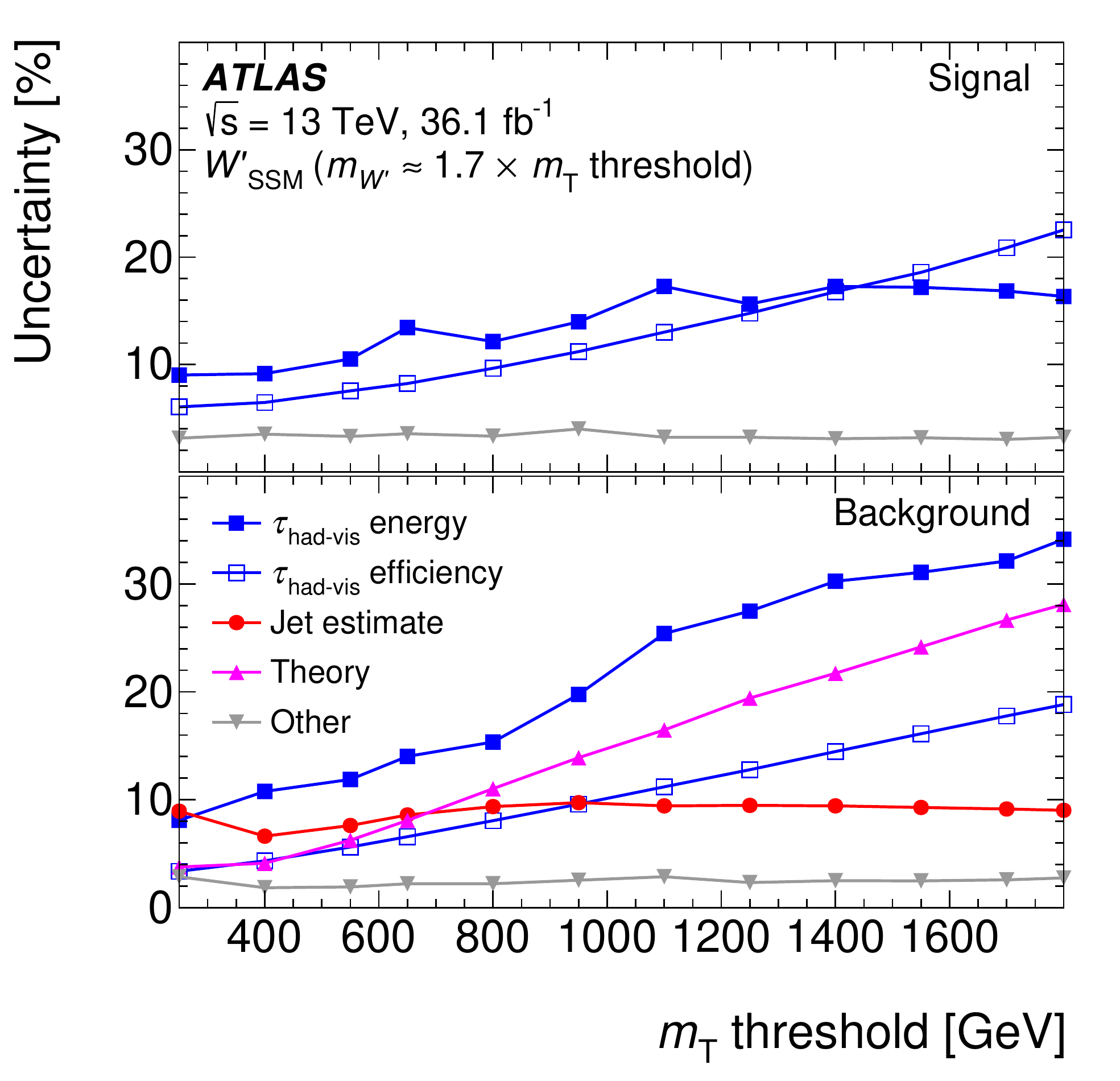}
    \caption{
        Dominant relative uncertainties in the expected signal and background
        contributions as a function of the \mt threshold. For each 
        threshold a \WprimeSSM boson with a mass of approximately 
        1.7 times the threshold is chosen.
        \emph{Theory} includes uncertainties in the cross sections
        used to normalize the simulated samples and uncertainties associated with the
        modeling provided by the event generators.
        \emph{Other} is the impact of all other uncertainties added in quadrature. 
     }
    \label{fig:systematics}
\end{figure}

Figure~\ref{fig:finalmass} shows the observed \mt{} distribution of the data
after event selection, including the estimated SM background contributions and
predictions for \WprimeSSM and \WprimeSFM ($\cot\phinu=5.5$) bosons with masses
of \SI{3}{\TeV}.  The number of observed events is consistent with the expected
SM background.  Therefore, upper limits are set on the production of a
high-mass resonance decaying to \taunu. The statistical analysis uses a
likelihood function constructed as the Poisson probability describing the total
number of observed events given the signal-plus-background expectation.
Systematic uncertainties in the expected number of events are incorporated into
the likelihood via nuisance parameters constrained by Gaussian prior
probability density distributions.  Correlations between signal and background
are taken into account.   A signal-strength parameter, with a uniform prior
probability density distribution, multiplies the expected signal.  The dominant
relative uncertainties in the expected signal and background contributions are
shown in Figure~\ref{fig:systematics} as a function of the \mt threshold.

\begin{figure}[tp]
    \centering
    \includegraphics[width=0.49\textwidth]{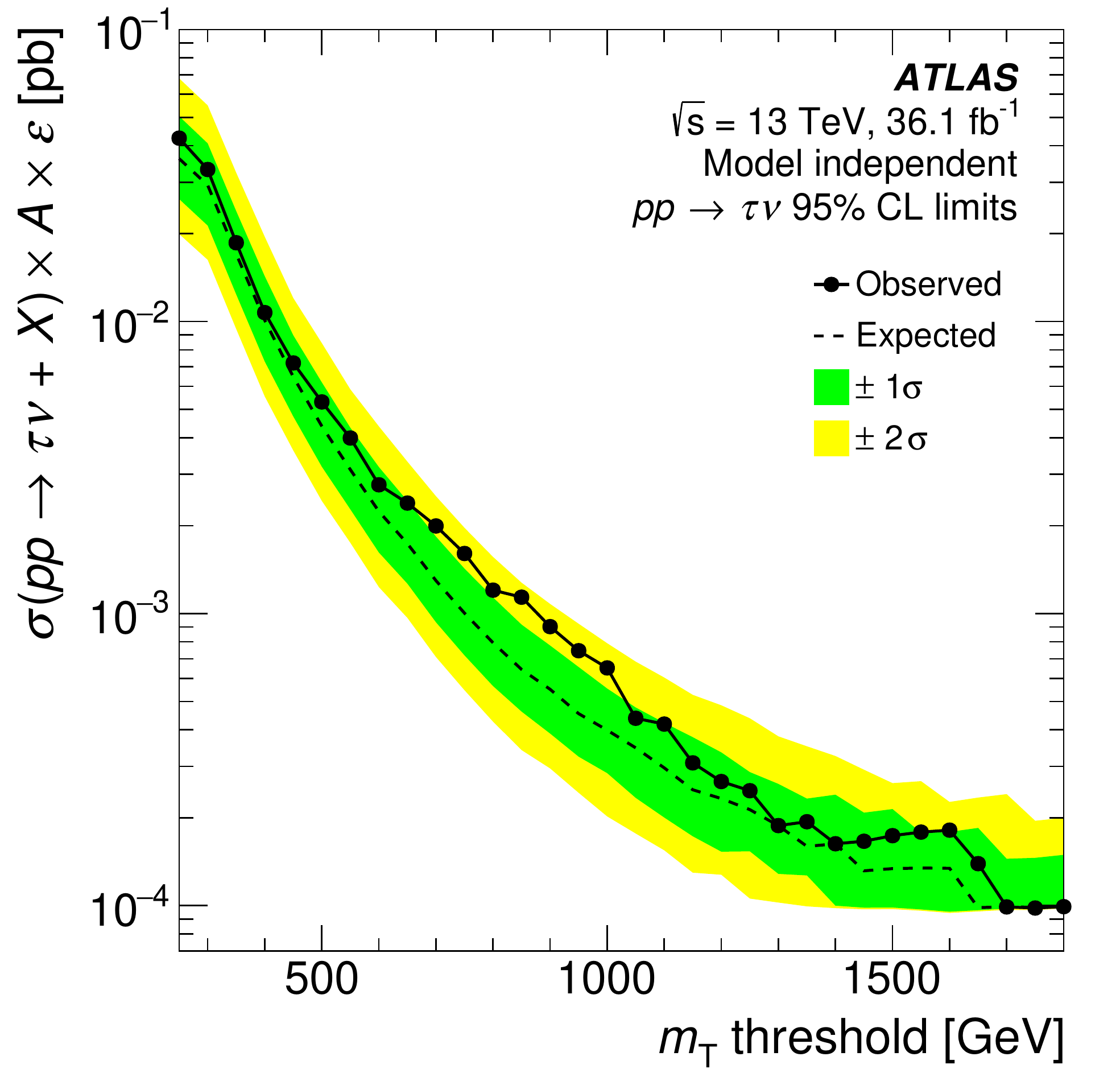}
    \caption{The 95\% CL upper limit on the visible $\tau\nu$ production 
        cross section as a function of the \mt threshold. 
    }
    \label{fig:fiducial}
\end{figure}

Limits are set at the 95\% credibility level (CL) using the Bayesian Analysis
Toolkit~\cite{BAT}. Figure~\ref{fig:fiducial} shows the model-independent upper
limits on the visible $\tau\nu$ production cross section, $\sigma(pp \to
\tau\nu +X)  \mathcal{A}  \varepsilon$, as a function of the \mt
threshold, where $\mathcal{A}$ is the fiducial acceptance (including the \mt
threshold) and $\varepsilon$ is the reconstruction efficiency. Model-specific
limits can be derived by evaluating $\sigma$, $\mathcal{A}$ and $\varepsilon$
for the model in question and checking if the corresponding visible cross
section is excluded at any \mt threshold. This allows the results to be
reinterpreted for a broad range of models, regardless of their \mt
distribution.  Good agreement between the generated and reconstructed \mt
distributions is found, indicating that a reliable calculation of the \mt
threshold acceptance can be made at generator level. The reconstruction
efficiency depends on \mt, $\varepsilon(\mt\,[\TeV]) = 0.633 - 0.313\mt +
0.0688\mt^2 - 0.00575\mt^3$, ranging from 60\% at \SI{0.2}{\TeV} to 7\% at
\SI{5}{\TeV}, and must be appropriately integrated out given the \mt
distribution of the model. The relative uncertainty in the parameterized
efficiency due to the choice of signal model is $\sim$10\%.  With these inputs
the visible cross sections for \WprimeSSM and \WprimeNU bosons could be
reproduced within 10\% using only generator-level information.  Data and
details to facilitate reinterpretations can be found at Ref.~\cite{HepData}. 

\begin{figure}[tp]
    \centering
    \subfigure[\label{fig:ssm_limit}]{\includegraphics[width=0.49\textwidth]{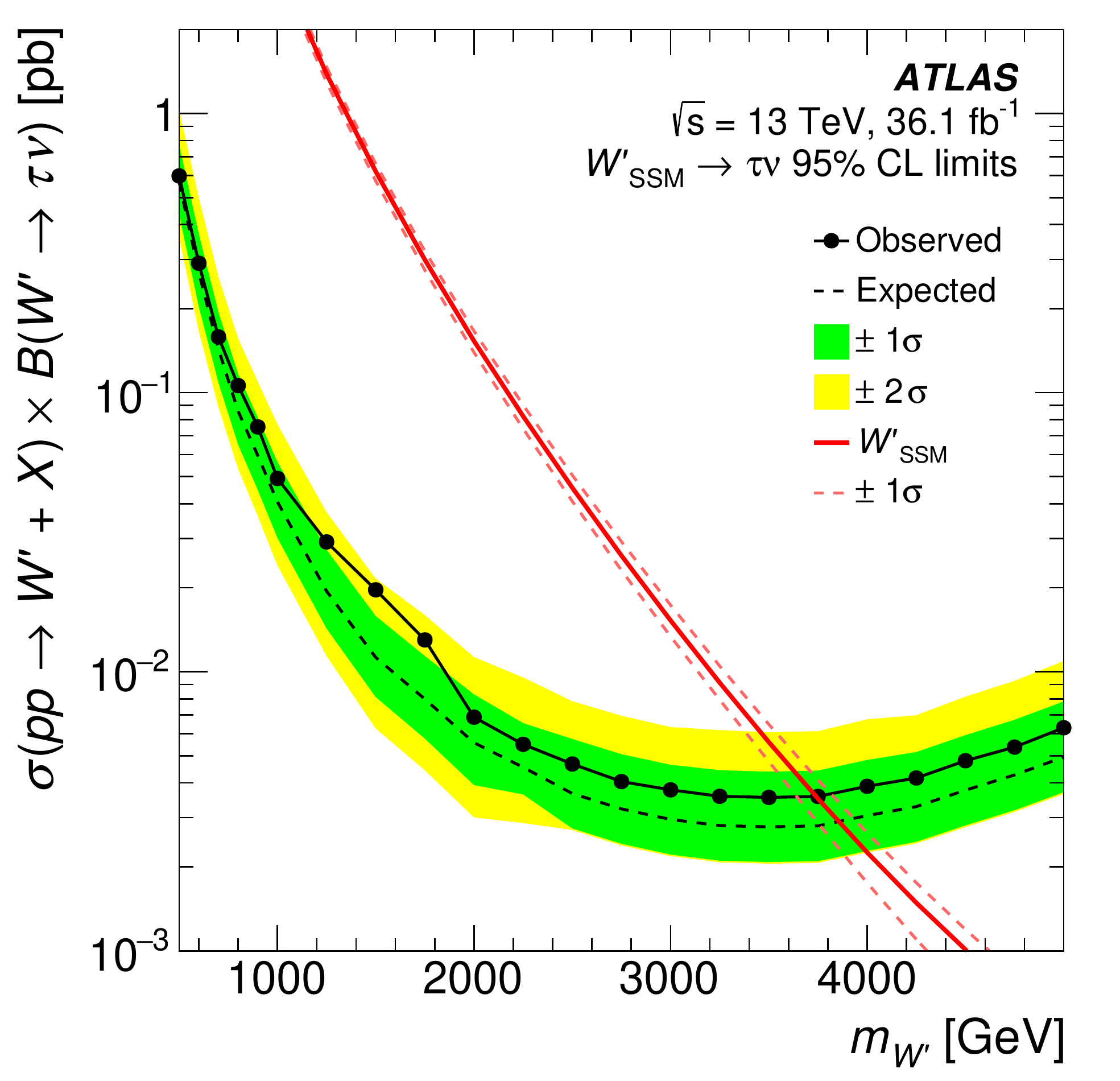}}
    \subfigure[\label{fig:nug_limit}]{\includegraphics[width=0.49\textwidth]{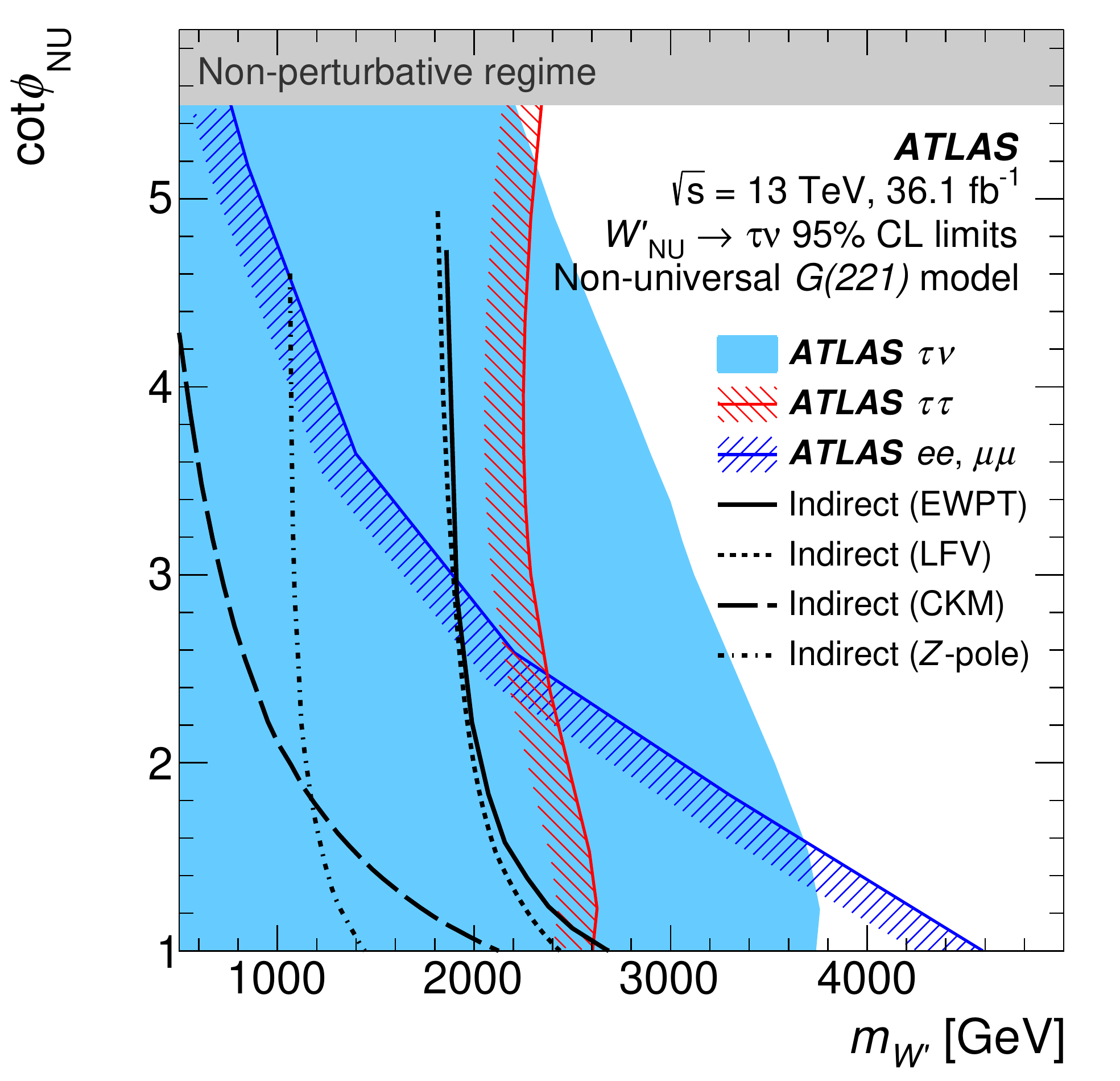}}
    \caption{
        (a) The 95\% CL upper limit on the cross section times \taunu branching fraction for \WprimeSSM.
        The \WprimeSSM cross section is overlaid where the additional lines 
        represent the total theoretical uncertainty.
        (b) Excluded region for \WprimeSFM.
        The 95\% CL limits from the ATLAS $ee$, $\mu\mu$~\cite{EXOT-2016-05} 
        and $\tau\tau$~\cite{HIGG-2016-12} searches         
        and indirect limits at 95\% CL from fits to electroweak
        precision measurements (EWPT)~\cite{G221LHC}, lepton flavor violation
        (LFV)~\cite{LeeLFV}, CKM unitarity~\cite{LeeCKM} and the original $Z$-pole
        data~\cite{sfm} are overlaid.
     }
    \label{fig:model_limit}
\end{figure}

Limits are also set on benchmark models by selecting the most sensitive \mt
threshold for each \Wprime mass hypothesis ($\sim$0.6\,$m_{\Wprime}$ up to a
maximum of \SI{1.45}{\TeV}).  The chosen threshold is found to have little
dependence on the \Wprime width. Figure~\ref{fig:ssm_limit} shows the 95\% CL
upper limit on the cross section times branching fraction as a function of
$m_{\Wprime}$ in the SSM. Heavy \WprimeSSM bosons with a mass lower than
\SI{3.7}{\TeV} are excluded, with an expected exclusion limit of
\SI{3.8}{\TeV}.  Figure~\ref{fig:nug_limit} shows the excluded region in the
parameter space of the non-universal $G(221)$ model.  Heavy \WprimeSFM bosons
with a mass lower than 2.2--\SI{3.8}{\TeV} are excluded depending on
$\cot\phinu$, thereby probing a significantly larger region of parameter space 
than previous searches~\cite{CMS-EXO-12-011}. The \WprimeSFM limits are typically
weaker than the \WprimeSSM limits as the increased \Wprime width yields lower
acceptances, while the enhancement in the decay rate cancels with the
suppression in the production via first- and second-generation quarks.  Limits
from the ATLAS $ee$, $\mu\mu$ and $\tau\tau$
searches~\cite{EXOT-2016-05,HIGG-2016-12} are also overlaid, showing that the
$\tau\nu$ search is complementary and extends the sensitivity over a large
fraction of the parameter space.  These  results suggest that the $\tau\nu$
searches should be considered when placing limits on non-universal extended
gauge groups, such as those seeking to explain lepton flavor violation in $B$
meson decays.

\FloatBarrier

In summary, a search for $\Wprime\to\tau\nu$ in \lumi of $pp$ collisions at
$\sqrt{s} = \SI{13}{\TeV}$ recorded by the ATLAS detector at the LHC is
presented.  The channel where the $\tau$ decays hadronically is analyzed
and no significant excess over the SM expectation is found. Upper limits are
set on the visible cross section for \taunu production, allowing
interpretation in a broad range of models.  Sequential standard model
\WprimeSSM bosons with masses less than \SI{3.7}{\TeV} are excluded at 95\%
CL, while non-universal $G(221)$ \WprimeSFM bosons with masses less than
2.2--\SI{3.8}{\TeV} are excluded depending on the model parameters.

\section*{Acknowledgments}



We thank CERN for the very successful operation of the LHC, as well as the
support staff from our institutions without whom ATLAS could not be
operated efficiently.

We acknowledge the support of ANPCyT, Argentina; YerPhI, Armenia; ARC, Australia; BMWFW and FWF, Austria; ANAS, Azerbaijan; SSTC, Belarus; CNPq and FAPESP, Brazil; NSERC, NRC and CFI, Canada; CERN; CONICYT, Chile; CAS, MOST and NSFC, China; COLCIENCIAS, Colombia; MSMT CR, MPO CR and VSC CR, Czech Republic; DNRF and DNSRC, Denmark; IN2P3-CNRS, CEA-DRF/IRFU, France; SRNSF, Georgia; BMBF, HGF, and MPG, Germany; GSRT, Greece; RGC, Hong Kong SAR, China; ISF, I-CORE and Benoziyo Center, Israel; INFN, Italy; MEXT and JSPS, Japan; CNRST, Morocco; NWO, Netherlands; RCN, Norway; MNiSW and NCN, Poland; FCT, Portugal; MNE/IFA, Romania; MES of Russia and NRC KI, Russian Federation; JINR; MESTD, Serbia; MSSR, Slovakia; ARRS and MIZ\v{S}, Slovenia; DST/NRF, South Africa; MINECO, Spain; SRC and Wallenberg Foundation, Sweden; SERI, SNSF and Cantons of Bern and Geneva, Switzerland; MOST, Taiwan; TAEK, Turkey; STFC, United Kingdom; DOE and NSF, United States of America. In addition, individual groups and members have received support from BCKDF, the Canada Council, CANARIE, CRC, Compute Canada, FQRNT, and the Ontario Innovation Trust, Canada; EPLANET, ERC, ERDF, FP7, Horizon 2020 and Marie Sk{\l}odowska-Curie Actions, European Union; Investissements d'Avenir Labex and Idex, ANR, R{\'e}gion Auvergne and Fondation Partager le Savoir, France; DFG and AvH Foundation, Germany; Herakleitos, Thales and Aristeia programmes co-financed by EU-ESF and the Greek NSRF; BSF, GIF and Minerva, Israel; BRF, Norway; CERCA Programme Generalitat de Catalunya, Generalitat Valenciana, Spain; the Royal Society and Leverhulme Trust, United Kingdom.

The crucial computing support from all WLCG partners is acknowledged gratefully, in particular from CERN, the ATLAS Tier-1 facilities at TRIUMF (Canada), NDGF (Denmark, Norway, Sweden), CC-IN2P3 (France), KIT/GridKA (Germany), INFN-CNAF (Italy), NL-T1 (Netherlands), PIC (Spain), ASGC (Taiwan), RAL (UK) and BNL (USA), the Tier-2 facilities worldwide and large non-WLCG resource providers. Major contributors of computing resources are listed in Ref.~\cite{ATL-GEN-PUB-2016-002}.

\printbibliography

\clearpage 
\begin{flushleft}
{\Large The ATLAS Collaboration}

\bigskip

M.~Aaboud$^\textrm{\scriptsize 137d}$,
G.~Aad$^\textrm{\scriptsize 88}$,
B.~Abbott$^\textrm{\scriptsize 115}$,
O.~Abdinov$^\textrm{\scriptsize 12}$$^{,*}$,
B.~Abeloos$^\textrm{\scriptsize 119}$,
S.H.~Abidi$^\textrm{\scriptsize 161}$,
O.S.~AbouZeid$^\textrm{\scriptsize 139}$,
N.L.~Abraham$^\textrm{\scriptsize 151}$,
H.~Abramowicz$^\textrm{\scriptsize 155}$,
H.~Abreu$^\textrm{\scriptsize 154}$
% ...
\bigskip
\\
% ...
$^{12}$ Institute of Physics, Azerbaijan Academy of Sciences, Baku, Azerbaijan\\
% ...
$^{88}$ CPPM, Aix-Marseille Universit{\'e} and CNRS/IN2P3, Marseille, France\\
% ...
$^{115}$ Homer L. Dodge Department of Physics and Astronomy, University of Oklahoma, Norman OK, United States of America\\
% ...
$^{119}$ LAL, Univ. Paris-Sud, CNRS/IN2P3, Universit{\'e} Paris-Saclay, Orsay, France\\
% ...
$^{137}$ $^{(a)}$ Facult{\'e} des Sciences Ain Chock, R{\'e}seau Universitaire de Physique des Hautes Energies - Universit{\'e} Hassan II, Casablanca; $^{(b)}$ Centre National de l'Energie des Sciences Techniques Nucleaires, Rabat; $^{(c)}$ Facult{\'e} des Sciences Semlalia, Universit{\'e} Cadi Ayyad, LPHEA-Marrakech; $^{(d)}$ Facult{\'e} des Sciences, Universit{\'e} Mohamed Premier and LPTPM, Oujda; $^{(e)}$ Facult{\'e} des sciences, Universit{\'e} Mohammed V, Rabat, Morocco\\
% ...
$^{139}$ Santa Cruz Institute for Particle Physics, University of California Santa Cruz, Santa Cruz CA, United States of America\\
% ...
$^{151}$ Department of Physics and Astronomy, University of Sussex, Brighton, United Kingdom\\
% ...
$^{154}$ Department of Physics, Technion: Israel Institute of Technology, Haifa, Israel\\
$^{155}$ Raymond and Beverly Sackler School of Physics and Astronomy, Tel Aviv University, Tel Aviv, Israel\\
% ...
$^{161}$ Department of Physics, University of Toronto, Toronto ON, Canada\\
% ...
$^{a}$ Also at Department of Physics, King's College London, London, United Kingdom\\
$^{b}$ Also at Institute of Physics, Azerbaijan Academy of Sciences, Baku, Azerbaijan\\
$^{c}$ Also at Novosibirsk State University, Novosibirsk, Russia\\
$^{d}$ Also at TRIUMF, Vancouver BC, Canada\\
$^{e}$ Also at Department of Physics and Astronomy, University of Louisville, Louisville, KY, United States of America\\
% ...
$^{*}$ Deceased
\end{flushleft}


\end{document}